\documentclass[12pt]{article}
\newcommand{\be}{\begin{equation}}
\newcommand{\ee}{\end{equation}}
\newcommand{\beas}{\begin{eqnarray*}}
\newcommand{\eeas}{\end{eqnarray*}}
\newcommand{\bea}{\begin{eqnarray}}
\newcommand{\eea}{\end{eqnarray}}
\newcommand{\ba}{\begin{array}}
\newcommand{\ea}{\end{array}}
\newcommand{\nn}{\nonumber}

\newcommand{\e}{\epsilon}
\newcommand{\al}{\alpha}
\newcommand{\om}{\omega}
\newcommand{\g}{\gamma}
\newcommand{\de}{\delta}
\newcommand{\la}{\lambda}
\newcommand{\si}{\sigma}

\begin{document}
\title{
{\bf General covariance violation and \\ 
the gravitational dark matter.        \\
II. Vector graviton}}
\author{Yu.\ F.\ Pirogov
\\[0.5ex]
\it Theory Division, 
Institute for High Energy Physics,  Protvino, \\
\it RU-142281 Moscow Region, Russia
}
\date{}
\maketitle
\abstract{\noindent 
The (four componenet) vector graviton contained in  metric, the scalar
component incorporated, is attributed to the violation of
the general covariance to the residual isoharmonic one. In
addition to the previously studied (singlet) scalar graviton, the
vector graviton  may constitute one more fraction of the gravitational
dark matter. The gravity interactions  of the vector graviton, as well
as its impact on the continuous medium are studied.} 

\section{Introduction}

In the preceding paper~\cite{Pirogov1}, the author put forward the
concept that the violation of the general covariance (GC) may serve as
a reason d'etre for the existence  of the dark matter~(DM) of the
gravitational origin (or v.v.). The case study for the (singlet)
scalar graviton, as a simplest representative of such a matter, was
worked  out. The consistency of the  theory  of such a graviton
considered as a part of the metric (in addition to the massless tensor
graviton) was assured there by the existence the residual unimodular
covariance~(UC). We  refer the reader to ref.~\cite{Pirogov1} for the
general discussion of the GC violation in the context of  the
gravitational DM and for the details of the UC case. 
The  GC violation  gets its  natural description  in
the framework of the affine Goldstone approach to
gravity developed in ref.~\cite{Pirogov2}, to which we refer the
reader as well.\footnote{For a short exposition of the approach, see
refs.~\cite{Pirogov3}, \cite{Pirogov4}.} In the present paper,  we
continue studying the GC violation for the next in complexity case
of the  (four component) vector graviton, the respective scalar
component incorporated. The consistency of the theory  is
assured in this case by  the residual isoharmonic covariance (see
later on). Section~2 of the paper is devoted to the gravity
interactions of the vector graviton. First, the theory  is constructed
in the distinguished background coordinates. Then, it is developed in
the arbitrary observer's coordinates. In Section~3, the impact of the
vector graviton on the continuous medium is studied.

\section{Gravity}

\subsection{Background coordinates}

\paragraph{Lagrangian}

Remind shortly the framework of the affine Goldstone approach to
gravity~\cite{Pirogov2} to be used in what follows  in describing
the GC violation. It is postulated in the approach that
there exists the physical, not just auxiliary, gravitational
background (continuum). Let $x^\mu$, $\mu,\nu,\dots=0,\dots,3$ be the
observer's coordinates of a point   in the space-time and let
$\bar\xi^\al=\bar\xi^\al(x^\mu)$, $\al,\beta,\dots =0,\dots,3$ be  the
background-attached coordinates of the point. The indices $\al$, etc.,
undergo the (global) affine transformations. Relative to the
observer's coordinate transformations, these indices are blind and can
be considered  as those just numerating the scalars.
The theory starts in the background coordinates, wherein all the
generic physical properties are already predestined. The
observer's coordinates only replicate these properties in the 
coordinate dependent fashion.
The background coordinates may be considered as an analogue
of the comoving coordinates for the continuous medium. The dynamical
variable for gravity in the background coordinates is the
metric~$\g_{\al\beta}(\bar\xi)$. The latter is, roughly speaking, the
square of the affine Goldstone boson which proves to be the original
field variable for the gravity.\footnote{For the spin-half partiacles,
the affine  Goldstone boson itself should be used instead  of its
square~\cite{Pirogov2}.}

Let us study the classical theory of the  metric and
the   matter  with the action
\be\label{L}
I=\int(L_{g}+ \Delta L_{g}
+ L_m )
\sqrt{-\g}\,d^4\bar\xi,
\ee
where $L_{g}$ and $\Delta L_{g}$ are the gravity Lagrangians, $L_m$
is the matter one and  $\g=\det \g_{\al\beta}$. By the very
construction, the action is to be invariant under the
(global) affine symmetry (AS). Nevertheless, some parts of
the Lagrangian may formally admit more wide sets of the coordinates.
As for the gravity,  $L_g$ is chosen so as 
to allow the arbitrary coordinates, possessing thus the GC.
$\Delta L_g$ is supposed to be restricted  to a subset of the general
coordinates, being  GC violating with some residual covariance.  As
for the matter, $L_m$ may in general violate the GC, too. 

Conventionally, take as $L_g$  the modified 
Einstein-Hilbert Lagrangian of the Ge\-neral Relativity (GR):
\be\label{EH}
L_g=- M_P^2 \Big(\frac{1}{2}R(\g_{\al\beta})-\Lambda\Big),
\ee
with $M_P$ being the Plank mass, $R$ being the Ricci scalar  and
$\Lambda$ being the cosmological constant. In principle, there is
conceivable  any generally covariant modification of the
Lagrangian~(\ref{EH}). 
As for the extra gravity  Lagrangian $\Delta L_g$, decompose
it generically in two terms:
\be\label{DL_g}
\Delta L_g= \Delta K_g -\Delta V_g,
\ee
with $\Delta K_g$ being the derivative kinetic term 
and $\Delta V_g$ being the derivativeless potential.
Consider these terms in turn.

\paragraph{Kinetic term}

Take $\Delta K_g$ in the lowest approximation as follows:
\be\label{kin}
\Delta K_g(\om^\al)=\frac{1}{2}\kappa^2\,\om\cdot \om,
\ee
with $\kappa$ being a constant with the dimension of
mass, $\vert \kappa\vert<M_P$. A priori, both  $\kappa^2 \ge 0$  and
$\kappa^2 < 0$ can be envisaged, corresponding to physical particles
or ghosts, respectively. Here and in what follows in this subsection
the notation for the dot product 
$\om\cdot \om= \om_\al \om^\al $ is understood with the metric
$\g_{\al\beta}$, until stated otherwise (and similarly for any two
vectors). In the above, the Lagrangian variable for the
vector graviton is defined as follows 
\be\label{Lvar}
\om^\al=\g^{\beta\g}\Gamma^\al{}_{\beta\g}
+k \g^{\al\g}\Gamma^\beta{}_{\beta\g},
\ee
 with 
\be\label{Gamma}
\Gamma^\al{}_{\beta\g}=
\frac{1}{2}\,\g^{\al\de}\Big(\partial_{\beta}
\g_{\de\g}+\partial_{\g}\g_{\de\beta}
-\partial_\de \g_{\beta\g}\Big)
\ee
being the Christoffel connection  and $k$  any real.  Eq.~(\ref{Lvar})
is the most general expression for the  gravity vector variable
allowed by the AS~\cite{Pirogov2}. Remind that 
\bea\label{rem}
\g^{\beta\g}\Gamma^\al{}_{\beta\g}&=&-\frac{1}{\sqrt{-\g}}\,
\partial_\de(\sqrt{-\g}\g^{\al\de}),\nn\\
\Gamma^\beta{}_{\beta\g}&=&\partial_\g\ln \sqrt{-\g}
\eea
and thus
\be\label{omega}
\omega^\al=-\partial_\beta
\g^{\al\beta}+(k-1)\partial^\al
\ln\sqrt{-\g},
\ee
where $\partial^\al=\g^{\al\beta}\partial_\beta$.

\paragraph{Isoharmonic covariance}

Consider the arbitrary  change of the background coordinates:
$\bar\xi^\al\to\bar\xi'^\al=\bar\xi^\al+\bar\varepsilon^\al$. Under
this coordinate transformation, one has
\bea
\Gamma'^\al{}_{\beta\g}(\bar\xi')&=& \partial'_\beta\bar\xi^\de
\partial'_\g \bar\xi^\varepsilon\Big( \partial_\varphi\bar\xi'^\al
\Gamma^\varphi{}_{\de\varepsilon}(\bar\xi)
-\partial_\de\partial_\e\bar\xi'^\al\Big),\nn\\
\g_{\beta\g}(\bar\xi)&=&\partial_\beta\bar\xi'^\de
\partial_\g \bar\xi'^\varepsilon  \g'_{\de\varepsilon}(\bar\xi'),
\eea
where $\partial'_\al=\partial/\partial \bar\xi'^\al$, etc. This gives
for the small $\bar\varepsilon^\al$:
\be
{\om}'^\al(\bar\xi')= \om^\al(\bar\xi) +\om\cdot \partial
\bar\varepsilon^\al
-(\partial\cdot\partial \bar
\varepsilon^\al+k\partial^\al\partial\cdot\bar\varepsilon),
\ee
where $\om\cdot\partial=\om^\beta\partial_\beta$,
$\partial\cdot\bar\varepsilon=\partial_\beta\bar\varepsilon^\beta$ and
$\partial\cdot\partial=\partial^\beta\partial_\beta=\g^{\al\beta}
\partial_\al\partial_\beta$.
The similar notations for the dot product containing the (covariant)
derivatives,  mutatis mutandis, will be used in what
follows. For the theory to remain meaningful, only those background
coordinates are allowed under the substitution of which $\om^\al$
transforms homogeneously as a vector. This gives
\be\label{QHeq}
\partial\cdot\partial \bar
\varepsilon^\al+k\partial^\al\partial\cdot\bar\varepsilon=0.
\ee
Call this equation the isoharmonic one.\footnote{It is not to be
mixed with the conventional harmonicity condition,
$\g^{\beta\g}\Gamma^\al{}_{\beta\g}=0$.}   The respective
transformation group will be called the isoharmonic
one, resulting in  the isoharmonic covariance~(IC).
At $k=0$, eq.~(\ref{QHeq}) remains to be  valid for the
finite~$\bar\varepsilon^\al$. In the limit $|k|\gg 1$, it reduces (up
to a constant) to the unimodularity condition,
$\partial\cdot\bar\varepsilon=0$ (so that $\de \g=0$),
corresponding to the (singlet) scalar graviton~\cite{Pirogov1}.

\paragraph{Weak-field limit}

For the physics interpretation, consider the weak-field limit of the
theory corresponding to the decomposition 
\be
\g_{\al\beta}=\eta_{\al\beta}+ h_{\al\beta}, 
\ee
with $\vert h_{\al\beta}\vert \ll 1$. Accounting for
eq.~(\ref{omega}) and  $\g=-(1+h)$,  $h= \eta^{\al\beta}
h_{\al\beta}$, one gets
\be\label{00}
\om^\al=\partial_\beta \Big(h^{\al\beta}+\frac{1}{2}(k-1)
\eta^{\al\beta} h\Big). 
\ee
All the indices are manipulated in the weak-field limit by means of
$\eta_{\al\beta}$ and $\eta^{\al\beta}$. 
In the limit $|k|\gg1$, one gets $\om_\al\sim
\partial_\al h$ and thus this limit corresponds to the (singlet)
scalar graviton, indeed.

Consider the ``gauge'' transformations for the ``potentials''
$h_{\al\beta}$, initiated by the
small  transformations of the background coordinates:
\be
h'_{\al\beta}(\bar\xi)=h_{\al\beta}(\bar\xi)-(\partial_\al\bar
\varepsilon_\beta
+\partial_\beta\bar\varepsilon_\al)
\ee
(and thus $h'=h-2\partial\cdot\bar\varepsilon$).
Under these gauge transformations, the field strength changes as 
\be
\om'^\al(\bar\xi)=\om^\al(\bar\xi)-(\partial\cdot\partial
\bar\varepsilon^\al
+k \partial^\al \partial\cdot\bar\varepsilon).
\ee
Clearly, $\om^\al$ is gauge invariant only under the isoharmonic
transformations eq.~(\ref{QHeq}), with the dot product defined now by
$\eta_{\al\beta}$. 

In the linearized GR in the flat background, the requirement for the
r.h.s.\ of eq.~(\ref{00}) to be zero is nothing but  the
Hilbert-Lorentz gauge condition eliminating from $h_{\al\beta}$ the
three vector and one scalar degrees of freedom. On the contrary,
aban\-doning this requirement in the present paper puts the respective
degrees of freedom in action. 
It is for this reason, that we interpret  the extra gravity components 
contained in $\om^\al$  as those corresponding to the
vector graviton, the scalar component incorporated. 
In this, $\om^\al$ is nothing but the field strength for such a
graviton. The existence of the residual IC is crucial for the theory.
It allows one to preserve the 
well-established properties of the  tensor gravity. Namely, 
the  IC serves as the gauge symmetry to remove from $h_{\al\beta}$ the
remaining (singlet) scalar and three tensor components, leaving
thus six physical components: four for the vector graviton, the scalar
component incorporated, and two for the massless tensor graviton. 

Note for completeness, that including in $\Delta K_g$ the term
quadratic in $\Gamma^\al{}_{\al\beta}$ with the independent
coefficient $\kappa_0^2$ produces additionally  the physical
(singlet) scalar graviton, followed by
the unimodularity condition, $\partial\cdot \bar\e=0$, and the
residual unimodular isoharmonic covariance (UIC). This  realizes
the most general case for the scalar and vector gravitons, described
by the three independent constants.

\paragraph{Mass term}

To attribute the mass to the extra graviton one should account
additionally for the  potential $\Delta V_g$.
The latter is a scalar function $\Delta V_g((\g\eta)_\al{}^\beta)$
depending on the determinant $\g$   
and $\mbox {\rm tr} ((\g\eta)_\al{}^\beta)^n$, with  $n$ being
an arbitrary integer, positive or negative, and $\eta^{\al\beta}$
being the Minkowski symbol.  
Clearly, using the latter  violates explicitly the AS to the Lorentz
one, which is supposed to be exact. In this, the general covariance is
violated completely.  The degree of this
violation  is characterized by a  mass parameter~$\mu$. 
Thus in the framework of  the affine Goldstone approach to gravity,
one  can structure the gravity Lagrangians, 
with $L_g(R)$ being both affine invariant and generally covariant,
$\Delta L_g(\om^\al)$ being also
affine invariant though  GC violating, whereas $\Delta V_g((\g
\eta)_\al{}^\beta)$ violating both the AS and the GC.
The AS being the basic one in the given framework, one
expects $M_P>|\kappa|\gg\mu$ with the natural hierarchy  of the
residual covariance groups 
\be
GC\stackrel{\kappa}{\longrightarrow}
IC\stackrel{\mu}{\longrightarrow} TC
\ee
for $L_g$, $\Delta K_g$ and $\Delta V_g$, respectively (TC meaning
the trivial covariance). Because the potential supplies  the mass to
the tensor graviton, too, we postpone the mass issue to the cumulative
study of the graviton mass mixing in the future.

Varying the action (\ref{L}) with respect to $\g^{\al\beta}$ one would
get the equation of motion for the gravity in the basic form. 
Then one could transform the results into the observer's coordinates.
For the physics generality, we rewrite the Lagrangian directly in the
observer's coordinates and proceed therein. The results in the
background coordinates will be recovered  as a marginal case.

\subsection{Observer's coordinates}

\paragraph{Lagrangian}

The action now looks like
\be\label{actionx}
I=\int(L_{g}+ \Delta L_{g}
+ L_m )
\sqrt{-g}\,d^4  x,
\ee
with  the metric 
\be
g_{\mu\nu}=\partial_\mu
\bar\xi^\al\partial_\nu\bar\xi^\beta
\g_{\al\beta},
\ee
(the inverse one $g^{\mu\nu}=\partial_\al
x^\mu\partial_\beta x^\nu \g^{\al\beta}$)   
and $g=\det g_{\mu\nu}$. Throughout  this subsection, all the indices
are manipulated by means of  $g_{\mu\nu}$ and $g^{\mu\nu}$, unless
stated otherwise. The generally covariant Lagrangian $L_g$ gets
unchanged:
\be\label{EH'}
L_g=- M_P^2 \Big(\frac{1}{2}R(g_{\mu\nu})-\Lambda\Big).
\ee
To proceed, introduce the auxiliary fields
\bea\label{aux}
\bar g_{\mu\nu}&=&\partial_\mu \bar\xi^\al\partial_\nu \bar\xi^\beta
\eta_{\al\beta},\nn\\
\bar g^{-1\mu\nu}&=&\partial_\al x^\mu \partial_\beta
x^\nu
\eta^{\al\beta}.
\eea
By definition, call  $\bar g_{\mu\nu}$  the background metric ($\bar
g^{-1\mu\nu}$ being the  inverse  one).\footnote{Note that $\bar
g^{-1\mu\nu}$ is to be distinguished
from $\bar g^{\mu\nu}= g^{\mu\la} g^{\nu\rho}\bar
g_{\la\rho}$.}
The mass term is modified straightforwardly to the scalar function
$\Delta V_g((g \bar g^{-1})_\mu{}^\nu)$ which depends on the ratio of
the determinants, $g/\bar g$,  and 
$\mbox {\rm tr} ((g\bar g^{-1})_\mu{}^\nu)^n$. 
The matter Lagrangian $L_m$ will be
discussed in the next Section.

\paragraph{Kinetic term}

As for $\Delta K_g$, it gets modified as follows:
\be\label{om}
\om^\la =
\Omega_{}^\la-\bar\Omega_{}^\la ,
\ee
where 
\bea\label{Om''}
\Omega_{}^\la &=& g^{\mu\nu}\Gamma^\la{}_{\mu\nu}
+k g^{\la\nu} \Gamma^\mu{}_{\mu\nu},
\nn\\
\bar\Omega_{}^\la &=& g^{\mu\nu}\bar\Gamma^\la{}_{\mu\nu}
+k g^{\la\nu}\bar\Gamma^\mu{}_{\mu\nu}.
\eea
In the above, $\Gamma^\lambda{}_{\mu\nu}$
is  the dynamical Christoffel connection defined, mutatis mutandis, by
eq.~(\ref{Gamma}) through $g_{\mu\nu}$. 
Accounting for the reduced connections
$g^{\mu\nu}\Gamma^\la{}_{\mu\nu}$ and
$\Gamma^\mu{}_{\mu\nu}$, given by eq.~(\ref{rem}) with the proper
substitutions, one gets similarly to eq.~(\ref{omega}):
\be\label{Om'}
\Omega^\la=-\partial_\nu g^{\la\nu}+(k-1)\partial^\la
\ln\sqrt{-g}.
\ee
By the construction, the symbol $\bar\Gamma^\lambda{}_{\mu\nu}$ is as
follows
\be\label{bc}
\bar\Gamma^\lambda{}_{\mu\nu}=\partial_\al x^\la
\partial_\mu\partial_\nu \bar\xi^\al.
\ee
It can  be expressed in terms of $\bar g_{\mu\nu}$ as the respective
Christoffel connection
\be\label{BarCh}
\bar\Gamma^\lambda{}_{\mu\nu}=
\frac{1}{2}\,\bar g^{-1\lambda\rho}\Big(\partial_{\mu}\bar 
g_{\rho\nu}+ \partial_{\nu}\bar g_{\rho\mu}
-\partial_{\rho}\bar g_{\mu\nu}\Big),
\ee
so that one has, in particular,
\be
\bar\Gamma^\mu{}_{\mu\nu}=\partial_\nu\ln\sqrt{-\bar g}.
\ee

Thus ultimately, the gravity is described by the  fourteen independent
fields: ten  for the tensor $g_{\mu\nu}$ and four
for the scalars  $\bar\xi^\al$. The latter ones are to be found
through observations, with the former ones being 
reproduced by the dynamics in the self-consistent fashion.  More
particularly, only the quadratic
combination of $\partial_\mu\bar\xi^\al$ in the form of $\bar
g_{\mu\nu}$, eq.~(\ref{aux}), enter.\footnote{For the spin-half
particles,  $\partial_\mu\bar\xi^\al$ themselves are operative.}
Due to the compensating terms with
the background connection, the theory can be studied in the
arbitrary observer's coordinates, in contrast to the background
coordinates. It should be stressed  that $\bar
g_{\mu\nu}$ (and thus $\bar\Gamma^\lambda{}_{\mu\nu}$)
is not the most general one, but depends on  the four scalar
parameter-fields $\bar\xi^\al$. 
It is for this reason, that there can be chosen the
coordinates where $\bar\Gamma^\la{}_{\mu\nu}=0$. From the
observer viewpoint, the last property  is precisely what
distinguishes the background coordinates from the remaining ones,
all of the coordinates being for the observer a priori equivalent.

\paragraph{Isoharmonic covariance}

It follows from eqs.~(\ref{om}), (\ref{Om''}) that $\om^\la$ is the
vector transforming homogeneously under the arbitrary change of the
coordinates  $x^\mu\to x'^\mu=x^\mu+\e^\mu$. 
This allows $\om^\la$ to serve as the Lagrangian
variable. The infinitesimal changes of the background and observer's
coordinates being related as
$\bar\varepsilon^\al=\partial_\mu\bar\xi^\al \e^\mu$,
the residual covariance group corresponds now to those $\e^\mu$ which
are related with $\bar\e^\al$,  satisfying eq.~(\ref{QHeq}). This
results
straightforwardly in the modified isoharmonic equation:
\be\label{Om}
\bar\nabla\cdot\bar\nabla \e^\la+k\bar\nabla^\la\bar\nabla\cdot\e=0.
\ee
Here $\bar\nabla_\mu$ is the  background covariant
derivative and
\bea\label{Om1}
\bar\nabla\cdot\bar\nabla&=&
g^{\mu\nu}\bar\nabla_\mu\bar\nabla_\mu,\nn\\
\bar\nabla^\la\bar\nabla\cdot\e&=& g^{\la\mu}
\partial_\mu\Big(\frac{1}{\sqrt{-\bar g}}\partial_\nu({\sqrt{-\bar
g}}\e^\nu)\Big). 
\eea
At $\bar\Gamma^\la{}_{\mu\nu}=0$, eq.~(\ref{Om}) reduces,
mutatis mutandis, to eq.~(\ref{QHeq}). In the limit $|k|\gg 1$, it
reduces to the modified unimodularity  condition
$\bar\nabla\cdot\e=0$ (up to a constant), or otherwise
$\partial\cdot(\sqrt{-\bar g}\e)=0$.

\paragraph{Weak-field limit}

In the observer's coordinates, the weak-field decomposition becomes
\be
g_{\mu\nu}=\bar g_{\mu\nu}+h_{\mu\nu}
\ee 
with  $h_{\mu\nu}=\partial_\mu\bar\xi^\al
\partial_\nu\bar\xi^\beta h_{\al\beta}$,  
$\vert h_{\mu\nu}\vert\ll 1$.
From eqs.~(\ref{Om''}) and (\ref{Om'}) with account for 
$\de \bar g_{\mu\nu}=0$,  one gets
\bea\label{deOm}
\de\Omega^\la&=&-\partial_\nu\de
g^{\la\nu}+(k-1)g^{\la\nu}\partial_\nu\de \ln\sqrt{-g}
\nn\\
&&+ (k-1)\partial_\nu\ln\sqrt{-g}\,\de g^{\la\nu},\nn\\
\de\bar \Omega^\la&=&\bar\Gamma^\la{}_{\mu\nu}\de g^{\mu\nu}
+k \bar\Gamma^\mu{}_{\mu\nu}\de g^{\la\nu}.
\eea
Accounting for the   the relations 
$\de \sqrt{-g}=-1/2\, \sqrt{-g} g_{\mu\nu}\de g^{\mu\nu}$, $\delta
g^{\la\rho}=-g^{\la\mu}g^{\rho\nu}\delta g_{\mu\nu}$
and  substituting $\de g_{\mu\nu}\to h_{\mu\nu}$,
$g_{\mu\nu}\to \bar g_{\mu\nu}$, one gets from eqs.~(\ref{om}) --
(\ref{Om'})
\be\label{0} \om^\la=\bar\nabla_\mu
\Big(h^{\la\mu}+\frac{1}{2}(k-1)\bar g^{\la\mu} h\Big), 
\ee
which clearly generalizes eq.~(\ref{00}).
Here one puts $h^{\mu\nu}= \bar g^{\la\mu}
\bar g^{\rho\nu} h_{\mu\nu}$,  $h= \bar g^{\mu\nu} h_{\mu\nu}$.
In the weak-field limit, the indices are manipulated by means of $\bar
g_{\mu\nu}$ and its inverse $\bar g^{\mu\nu}$.\footnote{In this limit
$\bar g^{\mu\nu}=\bar g^{-1\mu\nu}$.} 
Remind that one has, in particular, 
$\bar\nabla_\la \bar g^{\mu\nu}=0$, etc. 

The gauge transformations for the potentials $h_{\mu\nu}$
initiated by the infinitesimal
transformations of the observer's coordinates, the dependence of $\bar
g_{\mu\nu}$ on the coordinates including,  now look like
\be
h'_{\mu\nu}(x)= h_{\mu\nu}(x)- (\bar\nabla _\mu\e_\nu +
\bar\nabla _\nu\e_\mu)
\ee
(and thus $h'=h-2\bar\nabla\cdot \e$). 
Respectively, the field strength $\om^\la$ changes under the gauge 
transformations  as 
\be
\om'^\la(x)=\om^\la(x)-(\bar\nabla\cdot \bar\nabla\e^\la
+k\bar\nabla^\la \bar\nabla\cdot \e).
\ee
where
\bea
\bar\nabla\cdot\bar\nabla&=&\frac{1}{\sqrt{-\bar g}}\partial_\mu
(\sqrt{-\bar g}\,\bar g^{\mu\nu}\partial_\nu),\nn\\
\bar\nabla^\la\bar\nabla\cdot\e&=& 
\partial^\la\Big(\frac{1}{\sqrt{-\bar g}}\partial\cdot({\sqrt{-\bar
g}}\e)\Big). 
\eea
Clearly, the requirement for $\om^\la$ to be gauge invariant,
$\om'^\la(x)=\om^\la(x)$, results in the
isoharmonicity condition, eqs.~(\ref{Om}), (\ref{Om1}), with the
metric $g_{\mu\nu}$ substituted by~$\bar g_{\mu\nu}$.

\paragraph{Equations of motion}

Varying the action~(\ref{actionx}) with respect to $g^{\mu\nu}$ ($\bar
g_{\mu\nu}$ being unchanged)  one arrives at the gravity equation:
\be\label{eomg}
G_{\mu\nu}+ \Delta G_{\mu\nu} = M_P^{-2}\, T^{(m)}_{\mu\nu}.
\ee
Here   $G_{\mu\nu}$ is the gravity tensor defined as usually:
\be\label{GT}
-M_P^{2} G_{\mu\nu}
=\frac{2}{\sqrt{-g}}\,
\frac{\hspace{-1ex}\de {\cal L}_g}{\de
g^{\mu\nu}},
\ee
with ${\cal L}_g={\sqrt{-g}}\, L_g$ being the gravity Lagrangian
density (and similarly for $\Delta G_{\mu\nu}$ corresponding to
$\Delta L_g$). $T^{(m)}_{\mu\nu}$ is the conventional energy-momentum
tensor of the matter defined  through  ${\cal L}_m={\sqrt{-g}}\, L_m$
by the r.h.s.\ of eq.~(\ref{GT}).
Introduce the notation $(\nabla\cdot G)^\nu=\nabla_\mu G^{\mu\nu}$,
etc, with $\nabla_\mu$ being the generally covariant derivative.
Due to the GC of $L_g$ and thus the relation $(\nabla\cdot G)^\mu=0$,
one gets from eq.~(\ref{eomg})
the modified conservation law for the matter:
\be\label{DM}
\Big(\nabla\cdot (T_m+ \Delta T)\Big)^{\mu}=0.
\ee
The extra term $\Delta T^{\mu\nu}\equiv - M_P^2 \Delta G^{\mu\nu}$ is
to be interpreted as the contribution of the vector graviton to 
 the DM. In other terms, the equation above can be written as 
\be\label{DM'}
(\nabla\cdot T_m)^{\mu}=Q^\mu,
\ee
where 
\be
Q^\mu=M_P^{2}(\nabla\cdot\Delta  G)^{\mu}
\ee
is  the  external force acting on the matter from the site of the
vector gravitons.

Varying   the  Einstein-Hilbert Lagrangian density~${\cal L}_g$  with
respect to $g^{\mu\nu}$ one gets the gravity tensor as usually
\be
G_{\mu\nu}=R_{\mu\nu}-\frac{1}{2}R\,g_{\mu\nu}+\Lambda
g_{\mu\nu}. 
\ee 
With account for eq.~(\ref{deOm}), one gets from $\Delta {\cal L}_g$
\bea\label{sg}
- M_P^2\Delta G_{\mu\nu}&=&
\kappa^2\Big(
k(\om_\mu\partial_\nu\ln\sqrt{g/\bar g}+
\om_\nu\partial_\mu\ln\sqrt{g/\bar g}\,)-
\om_{\mu}\om_{\nu}
-\frac{1}{2}\,\om\cdot \om\, g_{\mu\nu}\Big)\nn\\
&&+\kappa^2\Big((k-1)\nabla\cdot \om\,g_{\mu\nu}+ (\bar{\nabla}_\mu
\om_\nu +\bar{\nabla}_\nu \om_\mu) \Big)\nn\\
&&+\Delta V_g\, g_{\mu\nu} -2 \partial \Delta V_g/ \partial
g^{\mu\nu}, 
\eea
where
\be\label{covdir}
\nabla\cdot \om=\frac{1}{\sqrt{-g}}\partial\cdot(\sqrt{-g}\,\om).
\ee
In the weak-field limit, $\bar\nabla_\mu$ coincides with $\nabla_\mu$.
In the limit $|k|\gg 1$  with the 
re\-definition  $\kappa\to \kappa/k$,  one recovers the
results for the (singlet) scalar graviton~\cite{Pirogov1}. 
At $\bar\Gamma^\la{}_{\mu\nu} =0$, mutatis
mutandis, the results in the background coordinates follow.

\section{Matter}

\paragraph{Energy-momentum tensor}

For the applications to cosmology, it suffices to treat the matter as
the continuous medium. Describe it directly in the observer's
coordinates with the metric $g_{\mu\nu}$. To apply the
Lagrangian framework to the medium~\cite{Pirogov1},  characterize the
latter by the  proper (i.e., measured in the comoving coordinates)
concentration $n$ of the medium particles  and by  the specific
entropy~$\si$  (the entropy per particle), in addition
to the medium 4-velocuty~$U^\mu$, $U\cdot U=1$.  
Besides, the medium is characterized by the nondynamical
parameters such as the particle mass~$m$, etc.  The
particle  number current $N^\mu=n U^\mu$
satisfies the generic continuity condition $\nabla\cdot N=0$.
This constraint has to be  valid identically, independently of the
equations of motion. 

Take the Lagrangian for the  medium generically as
\be\label{P}
L_m(n,\si, U^\mu)= - E (|N|,\si),
\ee
with the  scalar $E (|N|,\si)$ being  the energy function. Here
one puts $|N|= (N\cdot N)^{1/2}$.
One can also add  the  interactions of the vector
graviton with the medium  as follows:
\be
L'_{m}(n,\si,U^\mu)=-  F(|N|,\si)\frac{1}{|N|}\, N\cdot \om,
\ee
with the scalar $F(|N|,\si)$ being  the formfactor.
Introducing   the vector density ${\cal N}^\mu=\sqrt{-g}\,N^\mu$ as
the independent variable and the respective 
scalar density $|{\cal N}|=({\cal N} \cdot
{\cal N})^{1/2}=\sqrt{-g}\,|N|$, wright the  
total  Lagrangian density  as follows:
\be\label{TLd}
{\cal  L}_m^{(tot)}  =
-\sqrt{-g}\bigg( E \Big(\frac{|{\cal N}|}{\sqrt{-g}}\,,
\si\Big)+F \Big(\frac{|{\cal N}|}{\sqrt{-g}}\, ,
\si\Big) \frac{1}{|{\cal N}|}\, {\cal N}\cdot \om\bigg)
+\lambda\,
\partial\cdot {\cal N}, 
\ee
with $\lambda$ being the Lagrange's multiplier.

Varying eq.~(\ref{TLd})  relative to $\lambda$ one
reproduces the continuity condition, $\partial\cdot{\cal N}=0$.
Varying ${\cal  L}_m^{(tot)}$ with respect to $g^{\mu\nu}$ and
accounting for $\de {\cal N}^\mu=0$, 
$\de (|{\cal N}|/\sqrt{-g})=
n/2\,(g_{\mu\nu}-U_\mu U_\nu)\de g^{\mu\nu}$ 
one gets the energy-momentum tensor for the matter 
\be\label{T}
T^{(m)}_{\mu\nu}=\rho U_\mu U_\nu+ p(U_\mu U_\nu-g_{\mu\nu})
+{\tau}_{\mu\nu},
\ee
with
\bea\label{rhop}
\rho&=&e -(k-1){\nabla}\cdot(f U),\nn\\
p&=&  n e'-e +(n f'-f) U\cdot \om+(k-1){\nabla}\cdot(f U),\nn\\
{\tau}_{\mu\nu}&=& 
f(U_\mu \om_\nu + U_\nu \om_\mu)
-k f\Big(U_\mu\partial_\nu \ln\sqrt{g/\bar g}+ U_\nu\partial_\mu
\ln\sqrt{g/\bar g}\,\Big)
\nn\\
&&-\Big(\bar {\nabla}_\mu(f U_\nu)+ \bar{\nabla}_\mu(f U_\nu)\Big).     
\eea
Here $e (n,\si)\equiv E (|N|,\si)\vert_{|N|=n}$, $e'=\partial e
(n,\si)/\partial n$  and likewise for~$f$. 
In the equations above, $\rho$ is the scalar
coinciding with the energy  per unit proper 
volume,  $p$ is the scalar coinciding with the (isotropic) pressure,
while  $f $ is the new scalar state function. 
The terms proportional to~$f$ distort the medium. Being of the odd
degree in the medium velocity these terms reflect
the energy dissipation/pumping  for  the medium in the  vector
graviton environment. As a result, there appears in
$T^{(m)}_{\mu\nu}$ one more
independent tensor  structure ${\tau}_{\mu\nu}$ 
accounting, in particular, for the anisotropy of the medium.
In the limit $|k|\gg1$ with the  redefinition 
$f\to f/k$, eq.~(\ref{rhop}) reproduces the results for
the (singlet) scalar graviton~\cite{Pirogov1}, in particular,
$\tau_{\mu\nu}=0$. The trace of the energy-momentum
tensor  gets modified as
\be T_{m}{}^\mu_\mu=\rho-3p+\tau^\mu_\mu.
\ee

\paragraph{Equations of motion}

As the  equation of motion for the continuous medium, there serves 
the conservation condition in the external field,  eq.~(\ref{DM'}),
for the  matter energy-momentum tensor. This
equation can be divided into two parts. First, projecting it on the
streamlines by multiplying  on $U_\mu$ and accounting for $U\cdot
\nabla_\nu U=0$, one gets the
energy equation for the medium in the vector graviton field: 
\be\label{ee}
\nabla\cdot\Big((\rho+p )U\Big)- U\cdot\partial p
+U\cdot(\nabla \cdot {\tau})
=q .
\ee
The scalar $q=U\cdot Q$ on the r.h.s.\ of the equation above coincides
with the power $Q_0$ depositing in (dissipating from)  the medium per
unit proper volume due to interactions with the vector gravitons.
Second, restricting  eq.~(\ref{DM'})  by the
projector $\Pi_{\mu\nu}=g_{\mu\nu}-U_\mu U_\nu$, 
$(\Pi\cdot U)_\mu=(g_{\mu\nu}-U_\mu U_\nu)U^\nu=0$, on
the hypersurface orthogonal to the streamlines one
gets the modified Euler equation:
\be\label{Ee}
(\rho+p)U\cdot\nabla U_\mu +U\cdot\partial p\, U_\mu -
\partial_\mu p +(\Pi\cdot(\nabla \cdot {\tau}))_{\mu}
=Q_\mu- q U_\mu.
\ee 
When all the terms above, but the first one proportional to $\rho$,
are missing eq.~(\ref{Ee}) is nothing but the geodesic condition:
$U\cdot\nabla U_\mu=0$. Otherwise, it describes the deviation of the
flow from the geodesics due to the medium pressure and the influence
of the vector graviton field.

\section{Conclusion}

The classical theory of gravity with 
the  GC violation and the   residual  IC can consistently
be constructed. The theory  describes the vector graviton, the  scalar
component incorporated, as a part of the metric (in addition to the
massless tensor graviton). Similarly to the
previously studied (singlet) scalar graviton, the vector
graviton may constitute one more fraction of the gravitational~DM. The
case study for the tensor graviton, as the remaining part of the
gravitational DM, is to be given in the future.

\end{document}